\documentclass{appolb}
\usepackage{epsfig}
\usepackage{amssymb}
\usepackage{color}

\begin{document}

\title{World stock market: more sizeable trend reversal likely in
February/March 2010}

\author{Stanis{\l}aw Dro\.zd\.z$^{1,2}$, Pawe{\l} O\'swi\c ecimka$^{1}$
\address{$^1$Institute of Nuclear Physics, Polish Academy of Sciences,
PL--31-342 Krak\'ow, Poland \\
$^2$Faculty of Mathematics and Natural Sciences, University of Rzesz\'ow, PL--
35-310
Rzesz\'ow, Poland}}

\maketitle

\begin{abstract}

Based on our "finance-prediction-oriented" methodology which
involves such elements as log-periodic self-similarity, the
universal preferred scaling factor $\lambda \approx 2$, and allows
a phenomenon of the "super-bubble" we analyze the 2009 world stock
market (here represented by the S$\&$P500, Hang Seng and WIG)
development. We identify elements that indicate the third decade
of September 2009 as a time limit for the present bull market
phase which is thus to be followed by a significant correction. In
this context we also interpret the Chinese stock market index SSE.

The third decade of September 2009 was accompanied with a stock
market correction typically within the range of $4-5\%$ worldwide.
Taking into account the market patterns that followed the time of
delivering the previous scenario we present an updated scenario
whose critical time corresponds to October 28, 2009.

Assuming quite evident (as of November 12, 2009) termination of
the correction due to the above critical time we extend -
consistently with our methodology - the stock market forecasting
scenario. The corresponding expected S$\&$P500 future trend is
shown in Fig.~5 and it supports a potential average continuation
of increases to as far into the future as the turn of
February/March 2010. We also indicate the log-periodic patterns on
the gold market and they point to the end of November 2009 as the time when
the trend reversal - likely local however - is expected to begin.

\end{abstract}

Since the time - turn of February/March 2009 - of reaching the
deepest minima since 2003, the leading world stock market indices
started systematically elevating and, until present, they on
average increased within the range of $50 - 100\%$. Many other
world markets did follow this trend. In some cases this increase
was even stronger. Such a strong increase in a relatively short
period of time sooner or later has to terminate with a correction
that can be sizeable. A question that we address in this note is
whether this recent increase does reveal any precursors that
encode the date of this termination.  In the financial markets
such precursors are typically associated with the presence of
specific oscillations that get accelerated in time according to a
constant contraction factor $\lambda$ also referred to as a
preferred scaling factor. Based on many systematic analyzes and
our personal experience this factor corresponds to $\lambda
\approx 2$ if the rate of oscillatory contractions points to a
real reversal of the trend as postulated already
in~\cite{drozdz99,drozdz03} and further documented
in~\cite{bartolozzi}. As before~\cite{drozdz03,drozdz08}, for
transparency we use the simplest representation of the
corresponding log-periodic structure in the form
\begin{equation}
\Pi(\ln(x)/\ln(\lambda)) = A + B \cos({\omega \over 2\pi} \ln(x) + \phi),
\label{eq:FPE}
\end{equation}
where $\omega = 2\pi / \ln(\lambda)$. The stock market index
representation is then drawn according to the equation:
\begin{equation}
\Phi(x) = x^{\alpha} \Pi(\ln(x)/\ln(\lambda)), \label{eq:FP}
\end{equation}
where $x = \vert T - T_c \vert$ denotes a distance to the critical time $T_c$
and $T$ is
the clock time.

Three such current examples relating stock market indices
S$\&$P500, Hang Seng and WIG (Warsaw)- thus elected from entirely
different world zones - starting in February 1, 2009 and their
optimal log-periodic representations fulfilling our criteria of
consistency are shown in Figs.~1a, 1b and 1c respectively. These
three graphs were prepared on August 25, 2009 and then disclosed
on http://picasaweb.google.com/finpredict. They all three point to
the beginning of the third decade in September 2009 as a date
setting barrier for the present phase of increase.


\begin{figure}
\begin{center}

\includegraphics[width=0.6\textwidth,height=0.4\textwidth]{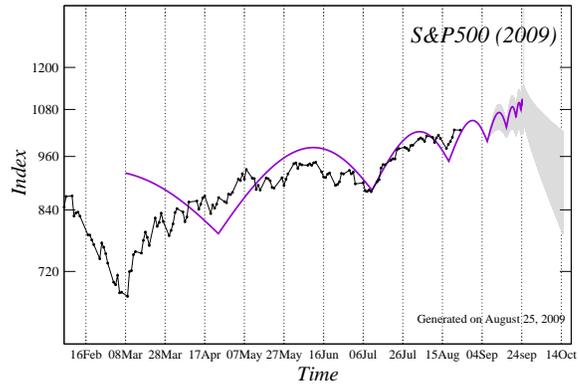}
\vspace*{1.5cm}

\includegraphics[width=0.6\textwidth,height=0.4\textwidth]{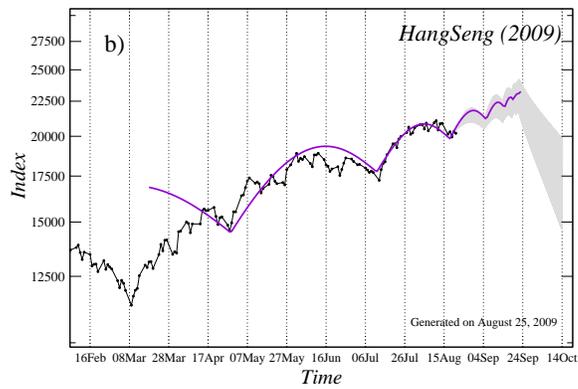}
\vspace*{1.5cm}

\includegraphics[width=0.6\textwidth,height=0.4\textwidth]{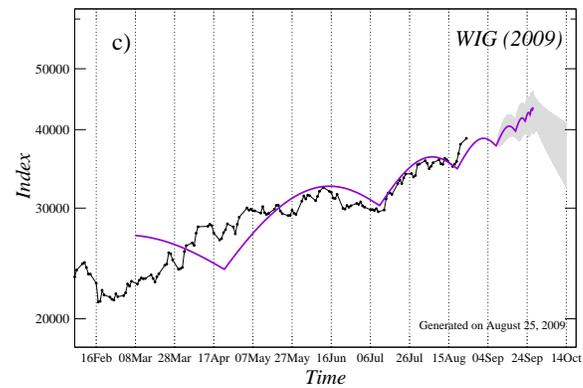}

\caption{The S$\&$P500 (a), Hang Seng (b) and WIG (c) stock market
indices versus their optimal ($\lambda = 2$) log-periodic
representations, prepared on August 25, 2009 and posted on
http://picasaweb.google.com/finpredict . The critical time
corresponds to the third decade in September 2009. The shaded
areas reflect current uncertainties regarding the specific course
of the development.} \label{fig1}
\end{center}
\end{figure}

Notice needs to be given here to the fact that we do not begin
tracking oscillatory patterns with the deepest minimum at the turn
of February/March 2009 but instead with a seemingly much less
convincing minimum in the second half of April 2009. One reason is
consistency which demands $\lambda \approx 2$. We also hypothesize
that this previous deepest minimum (here of February/March) is
still associated with the preceding market declining phase and
does not yet involve components characterizing the phase of
increase. That such a postulate is justified can be seen from
several historical stock market evolution examples. One such
example, particularly relevant in the present context, is shown in
Fig.~2. This is the S$\&$P500 development over the time period
2000 - present versus the corresponding log-periodic $\lambda = 2$
representations both decelerating and accelerating, depending on
the market phase. The up trend whose reversal took place in the
end of 2007 turns out to establish itself log periodically only in
the second half of 2004 and not already at the deepest levels in
2003. One should also notice that this last, by now historical,
example refers to a significantly larger time scale than currently
considered which provides further argument in favor of the concept
of log-periodic self-similarity~\cite{drozdz99,drozdz03}.


\begin{figure}
\begin{center}

\includegraphics[width=0.7\textwidth,height=0.5\textwidth]{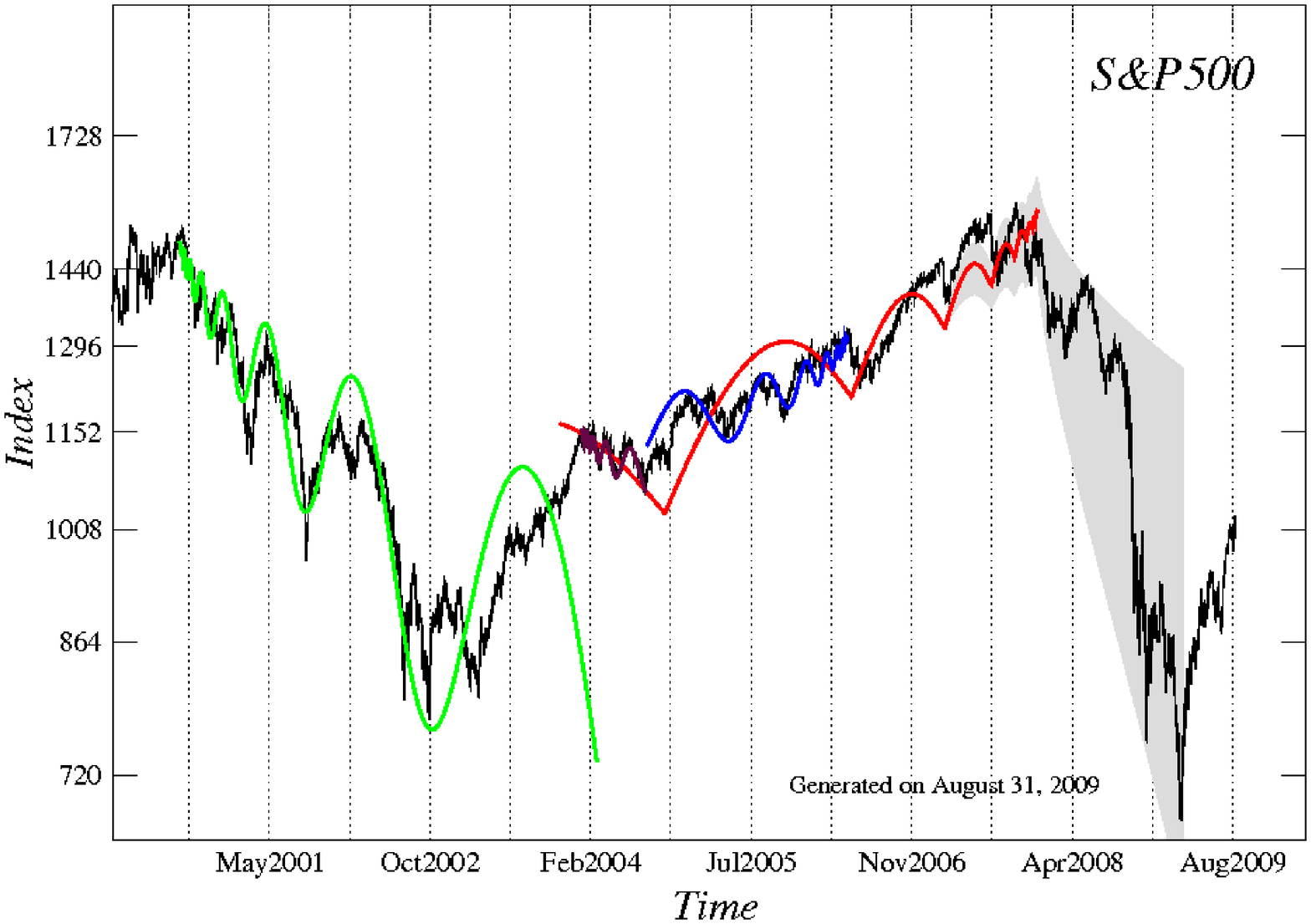}

\caption{The 2000 - present history of the S$\&$P500 stock market index
interpreted in
terms of the optimal log-periodic representations for different phases of its
development} \label{fig2}
\end{center}
\end{figure}

The above considered stock markets, similarly as many others in
the world, do correlate in phase since many years therefore their
oscillation patterns largely resemble each other and thus the
trend reversal is expected to occur at around the same time. There
is however one significant exception. This is the Chinese stock
market represented by the Shanghai SSE Composite Index. After
drawing down since October 2007, it started resuming the up trend
some two month earlier than the other world markets. As it can be
seen from today's perspective the end of this increasing phase
occurred on August 4, 2009 and this, as is shown in Fig.~3, can be
reproduced by the log-periodic function with the same contraction
factor $\lambda = 2$ but again by not taking into account the
initial absolute minimum. The other intermediate oscillations are
to be qualified as substructures corresponding to the shorter time
scales. Blindly taken relative magnitude of the amplitude of
oscillations operating on the different time scales may not always
directly reflect to what time scale a given pattern is to be
assigned. Some distortion, either artificial amplification or
reduction, may originate from some exogenous factors or, which is
especially likely in the present case, from some influence of the
other world markets on the Chinese market. This influence may of
course be mutual. It, in particular, may start pulling down the
other markets somewhat earlier~\cite {sornette03} than the above
predicted date specified as third decade in September 2009.


\begin{figure}
\begin{center}

\includegraphics[width=0.7\textwidth,height=0.5\textwidth]{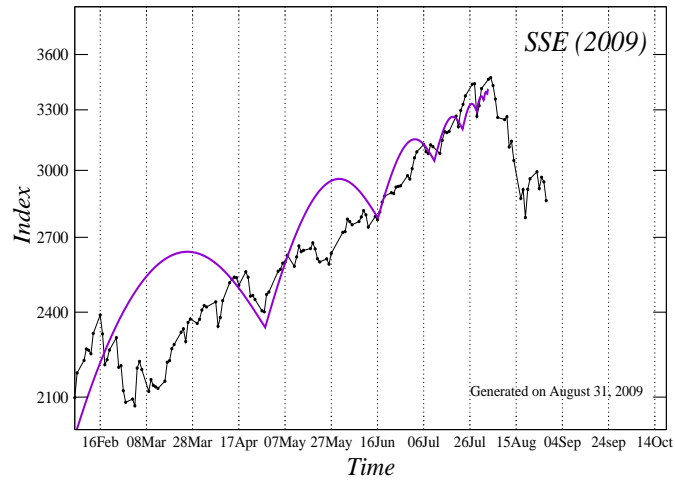}

\caption{The Shanghai Composite Index since February 1, 2009 until present
versus its
optimal log-periodic representation.} \label{fig3}
\end{center}
\end{figure}

(Note added on October 26, 2009)

In the third decade of September the world stock markets underwent
a correction within the range $4-5\%$ indeed but with some
exceptions (like Nikkei or Canadian S$\&$P TSX Composit)it
generally resumed the up trend and in October in most cases it
even topped the September maxima by 2-3$\%$. The question that we
address here is whether the additional market patterns developed
since the end of August 2009 allow to update (as compared to
Fig.~1) a consistent log-periodic scenario within the same
methodology as described above. The answer is presented in Fig.~4
for the same three stock market indices as before. Quality of the
resulting theoretical log-periodic market representation is
equally acceptable indeed and the critical time $T_c$ corresponds
to around October 28, 2009.


\begin{figure}
\begin{center}

\includegraphics[width=0.6\textwidth,height=0.4\textwidth]{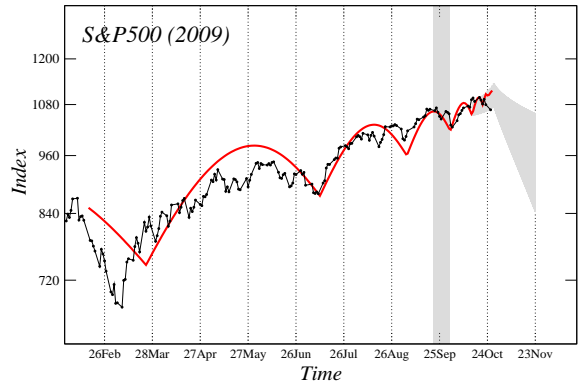}
\vspace*{1.5cm}

\includegraphics[width=0.6\textwidth,height=0.4\textwidth]{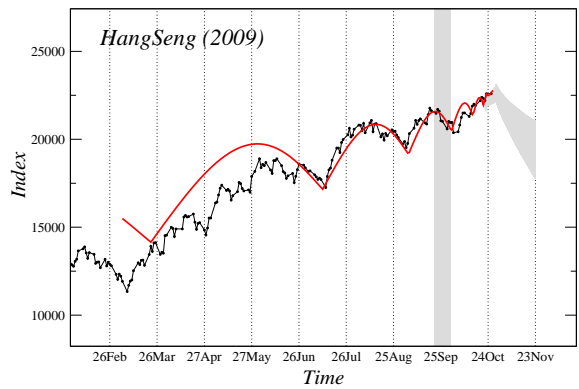}
\vspace*{1.5cm}

\includegraphics[width=0.6\textwidth,height=0.4\textwidth]{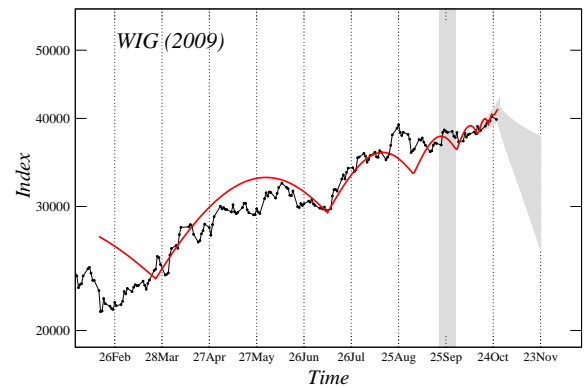}

\caption{The S$\&$P500 (a), Hang Seng (b) and WIG (c) stock market
indices versus their optimal ($\lambda = 2$) log-periodic
representations, prepared on October 26, 2009. The critical time
corresponds to October 28. The shaded vertical areas cover the
period of the third decade of September and the shaded areas after
October 28 reflect current uncertainties regarding the specific
course of the development during decline.} \label{fig4}
\end{center}
\end{figure}

A fundamental question that stays behind the markets dynamics is
this: suppose a law that governs it is identified and suppose more
and more market participants learn about it and start believing
the resulting predictions and thus act accordingly. If
sufficiently many participants (in an extreme case all) belong to
this group the law starts dissolving and eventually it disappears.
Perhaps its new variant starts emerging such that it remains
unperceived similarly as the previous one when it was not yet
commonly recognized. Empirical possibilities to verify such
effects by conducting realistic experiments on the real stock
market are of course very limited. Occasionally, under appropriate
conditions, the smaller local markets might offer a testing
ground. To some extent we in Poland seem to have experienced a
trace of action in this direction. Soon after we made or
prediction public (this in fact was one of our motivations),
accidentally or perhaps not entirely accidentally, there was a
growing expectation in a correction that is to come. As one
visible result, likely due to this fact, the Polish stock market
index WIG has been evolving sideways over the entire period of
September with an extremely strong tendency to the sudden sharp
drawdowns in response to even small moves down on the large world
markets where the global September trend was up. During the whole
month of September it did even not reach the top value of August.
Only after the third decade of September was over the situation
reversed and the Polish market become even more vital in its
tendency to move up. In any case the issue raised here remains an
intriguing subject for future studies, also in connection with the
concept of financial log-periodicity.

(Note added on November 16, 2009)

The critical time indicated before (turn of October/November) did
result in a correction of the order of $5-6\%$ (US, Asia)
up to $10\%$ (Europe), indeed. That however was not yet a longer
term trend reversal.  The previous as large correction as this
most recent one took place at the turn of June/July 2009. Based on
these facts and on the methodology described above, a natural
log-periodically extended scenario is shown in Fig.~5 (available
on \cite{finpredict} since November 12, 2009) for the S$\&$P500
(most leading world markets are likely to follow the same pattern
if applies). This scenario thus indicates that the stock market
carries potential to move up until around the turn of
February/March 2010.


\begin{figure}
\begin{center}

\includegraphics[width=0.7\textwidth,height=0.5\textwidth]{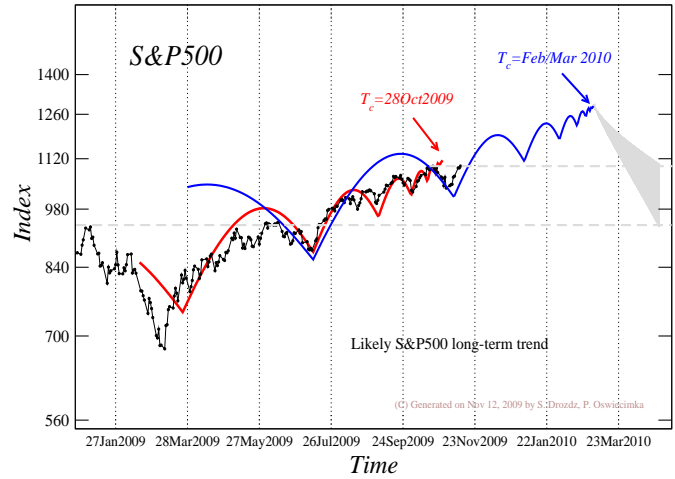}

\caption{The S$\&$P500 stock market index versus its optimal
($\lambda = 2$) log-periodic representation (prepared on November
12, 2009 and posted on \cite{finpredict}). The critical time
corresponds to the beginning of March 2010. The shaded area after
this period reflects current uncertainties regarding the specific
course of the development during decline.} \label{fig5}
\end{center}
\end{figure}

Concerning other markets viewed from similar perspective quite an
interesting development can be traced on the precious metals
market and especially on the gold market. The relevant
illustration is presented in Fig.~6 for the past one year starting
in October 2008. The related critical time corresponds to the
period just before the end of November 2009 which is to be
followed by a significant correction. This however does not yet
have to be an ultimate gold price trend reversal. The structure
under consideration is likely to constitute a subcomponent of the
longer-term up trend whose estimated termination would correspond
to the end 2010.

\begin{figure}
\begin{center}

\includegraphics[width=0.7\textwidth,height=0.5\textwidth]{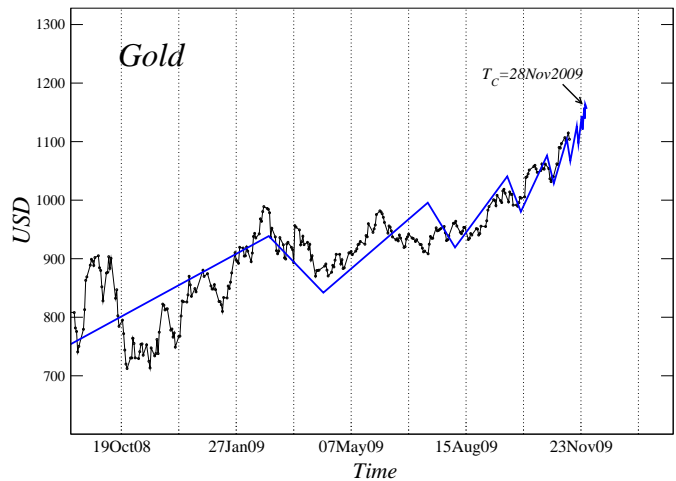}

\caption{The gold price over the period October 2008 until
November 12, 2009 versus its optimal ($\lambda = 2$) log-periodic
representation. The critical time corresponds to November 28,
2009} \label{fig6}
\end{center}
\end{figure}

\end{document}